\documentclass[preprint,12pt]{aastex}
\usepackage{epsfig}



\newcommand{\beq}{\begin{equation}}
\newcommand{\eeq}{\end{equation}}
\def\s-d{single-degenerate}
\def\d-d{double-degenerate}
\def\dd{double degenerate}

\def\ce{common envelope}
\def\rl{Roche lobe}
\def\mt{mass transfer}
\def\bis{binaries}
\def\ex{explosion}
\def\sn{supernova}
\def\sne{supernovae}
\def\nb{nebula}
\def\snb{supersoft nebula}
\def\dd{double degenerate} 

\def\snb{steady nuclear burning}

 \def\wdf{white dwarf}
 \def\sig{signature} 
 
   \def\Gc{Globular cluster}
\def\wdf{white dwarf} \def\t{temperature} 
\def\l{luminosity}

 \def\sss{supersoft source} 
\def\wdf{white dwarf}  
\def\sft{supersoft}   \def\sss{supersoft source}
\def\ne{nebula}       \def\pn{planetary nebula}
\def\te{temperature}  \def\lu{luminosit} \def\lus{luminosities} 
\def\sn{supernova}  \def\t1{Type Ia supernova} 
\def\ssn{supersoft nebula}  
\def\pr{progenitor}  \def\ac{accret} \def\bi{binary} \def\et{{\it et al.}}
\def\ev{evolution}
\def\per{$^{-1}$}
  
\def\three7{$10^{37}$~erg~s\per}  
\def\three6{$10^{36}$~erg~s\per}  
\def\three5{$10^{35}$~erg~s\per}  
\def\mt{mass transfer}
    
\def\rd{Di\thinspace Stefano} \def\sr{Rappaport} 
\def\pop{population} 
\def\lsxs{luminous supersoft X-ray source}

\def\bss{binary supersoft source} 
\def\mo{model}
\def\Rl{Roche lobe} 
 \def\s-b{steady-burning}
\def\nb{nuclear burning} 
\def\dy{dynamical}   
\def\ins{instability}

\def\do{donor} 
\def\s-d{single-degenerate} 
\def\D-d{Double-degenerate} 
\def\d-d{double-degenerate}

\def\dd{double degenerate} 
\def\N{NBWD} 
\def\Gc{Galactic}  
  
\def\obe{obscure} 
 
\def\rdtn{radiation}    
 \def\exn{explosion} 
\def\l-m{low-mass}
\def\h-m{high-mass}
\def\ch{{\it Chandra}} \def\xmm{{\it XMM-Newton}} \def\ro{{\it ROSAT}} 

\begin{document}

\title{
THE PROGENITORS OF TYPE IA SUPERNOVAE: I.~Are~they~Supersoft~Sources?}

\author{R. Di\thinspace Stefano
}
\affil{
Harvard-Smithsonian Center for Astrophysics,
60 Garden St..
Cambridge, MA 02138}

\begin{abstract}
In a canonical model, the progenitors of Type Ia 
supernovae (SNe Ia) are accreting, nuclear-burning
white dwarfs (NBWDs), which explode when the white dwarf (WD) reaches
the Chandrasekhar mass, $M_C.$ Such massive
NBWDs are hot ($k\, T \sim 100$ eV), luminous ($L \sim 10^{38}$ erg s$^{-1}$),
and are potentially observable as 
luminous supersoft
X-ray sources (SSSs). 
During the past several years, surveys for soft X-ray sources in external
galaxies have been conducted.
This paper shows that the results falsify the hypothesis
that a large fraction of progenitors are NBWDs which are 
presently observable as SSSs. The data also place limits on sub-$M_C$ models. 
While \t1\ \pr s may pass through one or more phases of SSS activity, these
phases are far shorter than the time needed to accrete most of the
matter that brings them close to $M_C.$ 
\end{abstract} 

\section{Introduction}

Type Ia supernovae (SNe Ia) help
us to explore the history of the cosmos (see, e.g.,
Riess et al.\, 2007; Kuznetsova et al.\, 2008). Yet we still know
little about their progenitors. 
There are two classes of progenitor models. (See Kotak 2009 and Branch
et al.\, 1995 for reviews.) 
In ``single degenerate" models, 
the progenitors are accreting white dwarfs (WDs). 
In ``double-degenerate'' models,  
the explosion is initiated through mass transfer between and/or the merger of
two WDs.
This paper is devoted to searching for the progenitors. 

In \S 2 we show that, if the single-degenerate model is dominant,
then galaxies such as our own and M31 contain 
on the order of $1000$ nuclear-burning
WDs (NBWDs) that will explode within $10^5$~years. 
In \S 3 we show that these progenitors could be hot and
bright enough to be detected 
as supersoft x-ray sources (SSSs) 
in many external galaxies. We then show that existing
data from {\it Chandra} place strict limits on the numbers of Type~Ia
progenitors that are detected as SSSs. In section \S 4 we
discuss the implications. These are explored more fully, in the context of
binary models for \t1\ \pr s, in the companion paper 
(Di\thinspace Stefano 2009).

\section{The Numbers of Accreting WD Progenitors}

In single-degenerate models, a WD must accrete
and retain matter. Let $M_0$ represent the initial mass
of the WD. An explosion occurs when the mass has increased to
a value $M_f.$ We start with models in which
$M_f$ is the Chandrasekhar mass, $M_C \approx 1.4\, M_\odot.$ 
The WD must have a donor that can contribute the requisite
mass. 
In addition, the rate of mass infall, $\dot M_{in}$,
 must be 
large enough that accreted matter can be burned and retained. 
The rate of  
genuine mass increase, $\dot M_{WD}$, is determined by the rate 
of infall to the WD and a retention factor 
$\beta(M_{WD},\dot M_{in})$.
\begin{equation}
\dot M_{WD} = \beta(M_{WD},\dot M_{in})\, \dot M_{in} 
\end{equation}
The top panel of Figure 1 illustrates that 
mass retention requires mass donation rates
in a narrow range between $\dot M_{in,min}(M_{WD})$ 
(roughly $10^{-7} M_\odot$~yr$^{-1}$ for a solar-mass CO WD),
and $\dot M_{in,max}(M_{WD})$, which is generally a few times 
larger than $\dot M_{in,min}(M_{WD})$.
This range corresponds to rates
of infall for which accreting matter can undergo 
quasisteady nuclear burning.\footnote{Note that even within the so-called
{\sl steady-burning region}, $\beta$ need not be equal to unity.
For example, some of the infalling mass may be driven from the 
\wdf\ in the form of winds.}  
For $\dot M_{in}>\dot M_{in,max}$, not all of the incoming matter can be 
burned, and $\beta$ declines. 
For $\dot M_{in}<\dot M_{in,min}$, incoming matter accumulates  
and burns episodically. 
For rates just below $\dot M_{in,min}(M_{WD}),$ the explosions are frequent
(separated by decades) and weak, corresponding to recurrent novae (RNe);
$\beta$ can be close to unity.  
The lower the rate of infall,
the more violent are the explosions produced by nuclear burning, and the
more matter is ejected from the system. 
In some cases, more matter may be ejected than is accumulated between novae; 
$\beta$ is negative. 
Most of the accreting WDs already identified are
cataclysmic variables with accretion rates typically two orders 
of magnitude or more below the rate required for steady burning.
Binaries can produce the requisite high rates, however
(van den Heuvel et al.\, 1992).

The duration, $\tau_{acc},$ of the accretion phase is 
\begin{equation}
\tau_{acc}=\int_{M_0}^{M_f} {{dM}\over{\dot M_{in}\, 
\beta(M_{WD},\dot M_{in})}} 
\end{equation}
There may be long periods in which the
mass transfer period is too low (or too  high) for mass retention.
In this paper we are concerned primarily with the interval during which
the value of $\dot M_{in}$ is large enough
that infalling mass is burned and retained. The approximate duration of this
interval can be
written as follows.   
\begin{equation}
\tau_{acc}=
 {{5}} \times 10^5\, yrs\,  
\Big({{\Delta\, M}\over{0.4\, M_\odot}}\Big)
\Big({{8\times 10^{-7} M_\odot\, yr^{-1}}\over{\beta\, \dot M_{in}}}\Big). 
\end{equation}
where the value of $\beta\, \dot M_{in}$ represents an average during
the epoch when the WD's mass is changing. 

The rate of SNe Ia in galaxies is roughly $0.3$
per century per $10^{10} L_\odot$ in blue luminosity
(Cappellaro et al.\, 1993; Turatto et al.\, 1994).
In a galaxy with blue luminosity $L_B,$ the number, $N_{acc}$, of
accreting progenitors within
 $0.4\, M_\odot$
of $M_C$ is roughly  
\begin{equation}
N_{acc}=
1500 \,  
\Bigg({{\Delta\, M}\over{0.4\, M_\odot}}\Bigg)  
\Bigg({{8\times 10^{-7} M_\odot\, yr^{-1}}\over{\beta \, \dot M_{in}}}\Bigg)  
\Bigg({{L_B}\over{10^{10} L_\odot}}\Bigg).
\end{equation}
To determine the expected range of values of $N_{acc},$ we  assess  
the range of likely values of each factor.  The amount of mass, 
$\Delta M,$  
that the \wdf\ must gain is constrained from below by the requirement that the 
accretor must start as a CO \wdf, while \wdf s born with the highest mass 
start as O-Ne-Mg \wdf s, which undergo accretion-induced collapse 
(AIC) upon
achieving $M_C$ (see, e.g., 
Nomoto et al.\, 1979).  The value of $\Delta M$
 can therefore not be smaller than 
$0.15-0.2\, M_\odot$.  
Furthermore, 
the numbers of high-mass \wdf s
are constrained by two considerations. First the 
initial mass function favors
low-mass stars (Miller \& Scalo 1979; Salpeter 1955). Second, low-mass 
stars produce low-mass \wdf s
(Kalirai et al. 2008; Catal{\'a}n et al. 2009 ; Dobbie et al. 2006; 
Williams 2007; Weideman \& Koester 1983).     
Therefore 
$\Delta M$ 
is almost certainly 
in the range:  $0.2-0.8\, M_\odot$, 
consistent with $N_{acc}$ in the range $750-2250$.  
In order for the average value of $\Delta M$ 
to be small enough to place $N_{acc}$ 
below $750$, the relationship between a star and the mass and/or 
composition of the remnant would have to be different from current models.  
Alternatively, 
part of the answer could be that O-Ne-Mg \wdf s produce 
\t1e. Either case would represent a shift in our understanding of 
fundamental astrophysics. 

If the average value of $\beta\, \dot M_{in}$ 
is larger than $8 \times 10^{-7} M_\odot$~yr$^{-1}$, 
this would reduce the value of $N_{acc}$.  
Higher rates of nuclear burning will, however, produce luminosities that 
exceed the Eddington luminosity.  It is more likely that the average rate at 
which mass is accreted is actually smaller, producing larger values 
of $N_{acc}$.  Note in addition that calculations by different groups agree 
on the general range of $\dot M_{in}$ needed for high-mass \wdf s 
to burn accreting matter (see, e.g., 
Nomoto 1982; Iben 1982, Fujimoto 1982; Shen \& Bildsten 
2008 and references therein).

The third factor influencing the value of $N_{acc}$ is proportional to $L_B.$
This factor is included because observations find that the  
rate of Type Ia supernova scales with $L_B.$   
As more \t1e are discovered in more galaxies, 
this rate is becoming better measured. (See, e.g.,
Dilday et al.\, 2008; Graham et al.\, 2008; Kuznetsova et al.\, 2008;  
Poznanski et al. 2008; Panagia et al.\, 2007.)  
It is unlikely that the estimated rates will 
change enough to reduce or 
enhance 
the predicted values of $N_{acc}$ by a factor as large as $2$.

Therefore,
taking into account the range of possible values of the 
factors in Equation 4, we find that 
for spiral galaxies similar to the Milky Way and M31 
(and even for ellipticals, with a lower rate of SNe Ia),
$N_{acc}$  cannot be much smaller than several hundred, and is likely to 
be significantly larger. 
The assumptions that lead to to these predictions are that (1)~the
\s-d\ channel produces the majority of \t1e, and (2) nuclear burning
of accreted material is needed to prevent explosive mass loss. 

\begin{figure*}
\begin{center}
\psfig{file=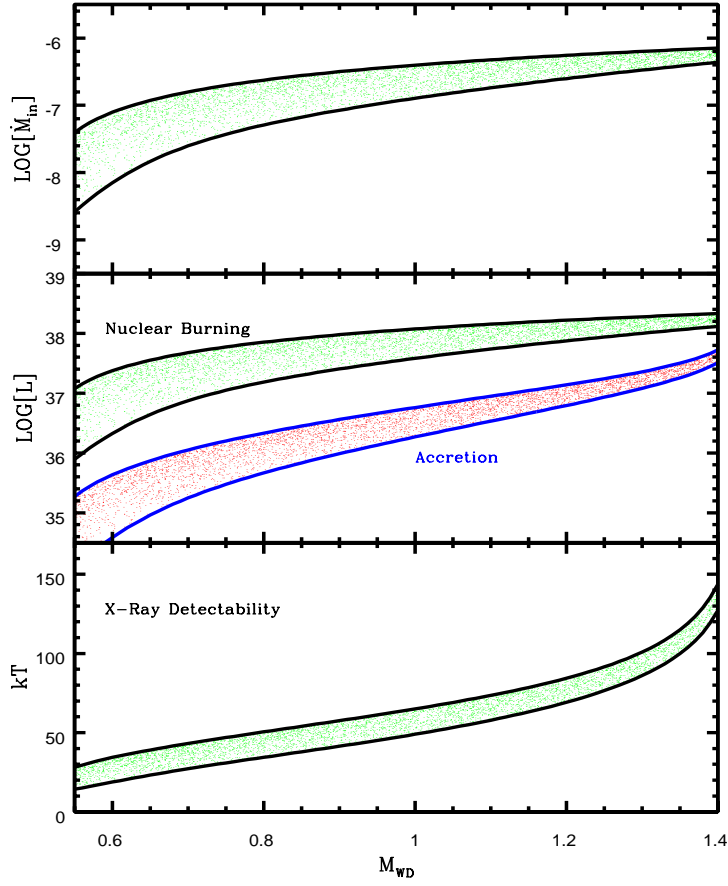,
height=5.0in,width=4.0in,angle=-0.0}
\caption{\footnotesize 
{\bf Properties of NBWDs:} 
green point correspond to systems in which the rate of mass infall
is in the range necessary for quasisteady nuclear burning.
In each panel, the green points lie between two curves;
the upper (lower) curves correspond to the maximum 
(minimum) rate of infall
consistent with quasisteady nuclear burning. 
{\bf Top panel:} $\dot M_{in}$ is plotted against the
white dwarf mass, $M_{WD}.$ For values of $\dot M_{in}$
below the steady-burning region, novae are expected. 
The repetition frequency is highest for $\dot M_{in}$ just below
the steady-burning limit, so that the novae are recurrent novae
and a large fraction of the accreted matter can be retained. For
even lower accretion rates, the novae are classical novae, blasting 
accreted material from the binary at intervals $> 10^4$~years.  
Just above the
steady burning region is the region in which mass comes in too
quickly for it all to be burned upon arrival.
Unless mass transfer can be turned off
during an interval in which all excess matter is burned, 
the value of $\beta$ must be smaller than unity.
{\bf Second panel:} The logarithm of the luminosity versus 
$M_{WD}$. The positions of the green points are 
computed by assuming that $0.75$ of the infalling matter
is burned with an efficiency of $0.007.$ The 
red points show the accretion luminosity, 
which 
was computed to be: $\dot M_{in} M_{WD} G/R_{WD}.$ 
Although more than an order of magnitude lower than the luminosity
derived from nuclear burning,
accretion by high-mass \wdf s is detectable in nearby external galaxies.
{\bf Third panel:} $k\, T$ versus $M_{WD}$. The value of 
$T$ was computed by assuming an effective radius equal to
$1.5\, R_{WD}$. This panel illustrates that the effective
temperature is a strong function of the WD mass. For WDs
with mass near $M_C,$ $k\, T$ is large enough to make
x-ray detection efficiency comparable to what it would be for
a canonical (i.e., hard) x-ray source. 
}
\end{center}
\end{figure*}

\section{Detecting the Progenitors} 

\subsection{The High Luminosity of Progenitors}  
The crucial point is that nuclear-burning, either episodic or quasisteady,
appears to be necessary for the retention of matter by WDs. 
The burning of matter releases a great deal of energy. 
\begin{equation} 
L_{NB} = 
2.4 \times 10^{38}~{\rm erg\, s}^{-1} 
\Bigg({{\eta}\over{0.007}}\Bigg)\, \Bigg({{f}\over{0.75}}\Bigg)\,  
\Bigg({{\beta\, \dot M_{in}}\over
{8 \times 10^{-7}M_\odot\, {\rm yr}^{-1}}}\Bigg)     
\end{equation} 
In this equation, $L_{NB}$ is the power released through the
burning of accreted material; $f$ is the fraction of the 
incoming matter burned,
and $\eta$ is the efficiency.   

Consider a WD with accretion rate in the steady-burning regime.
Because the mass infall rate is large, the accretion luminosity 
is also large, although typically less than $10\%$ of the
power provided by nuclear burning. If, however, the accretion rate
lies slightly below the steady-burning regime, then nuclear burning
will occur only episodically, during recurrent novae. In contrast, the 
accretion luminosity, which can be $\sim 10^{37}$~erg~s$^{-1}$
for high-mass \wdf s,
could make the system appear bright even between nova explosions.  
Therefore, particularly when nuclear burning is occurring, but 
even when mass infall rates are slightly below
 the nuclear-burning limit, accreting WDs that are Type~Ia progenitors 
should be highly luminous (second panel of Figure 1).  
We address the issue of the detectability of the accretion
luminosity in the companion paper (Di\thinspace Stefano 2009).
 
\subsection{Effective Temperatures: Detection at X-ray Wavelengths} 

 As the third panel of Figure 1 shows, effective values of $k\, T$ for NBWDs are expected
to be in the range of tens of eV. 
The x-ray missions {\it Einstein} and {\it ROSAT}
 had good low-energy sensitivity 
and could 
detect soft x-ray sources in the Local Group. 
By the mid-1990s, more than $30$ sources with the properties 
expected for NBWDs had been discovered. (See Greiner 2000 for a catalog and for 
references.) 
The soft emitters were dubbed 
luminous supersoft x-ray sources (SSSs).
A natural conjecture is that the SSSs are in fact NBWDs.
Some nearby SSSs are known to be
 hot white dwarfs post-nova, or are pre-WDs
still ensconced in
planetary nebulae. The binary properties of roughly half of the 
SSSs with optical IDs suggests that they might be WDs accreting at
high-enough rates to burn matter in a quasisteady fashion.
Soon after their discovery, it was conjectured that some SSSs could be
\pr s of AIC (van den Heuvel et al.\, 1992) or of \t1e (Rappaport, 
Di~Stefano \& Smith 1994). 

As Figure~1 demonstrates, the
temperature is higher for WDs with higher mass. This is
because higher-mass WDs (1)~are smaller, and (2)~require a higher infall
rate, and therefore a higher rate of nuclear burning.
In principle, NBWDs with mass close to $M_C$ can have values of 
$k\, T$ near or even slightly above $100$~eV. 

Di\thinspace Stefano \& Rappaport (1994)  
modeled the gas distribution of several galaxies to determine what
fraction of SSSs could be detected as a function of $L$ and $T$.
While only roughly a percent of low-L/low-T SSSs could be detected by 
{\it ROSAT}, almost all SSSs associated with NBWDs with mass near $M_C$
could be detected in M31, because they are both hot and bright.
Figure 2 demonstrates that, with {\it Chandra},
 for typical amounts of absorption, we 
can expect to have a complete census of NBWDs with masses within a few
tenths of $M_C$ in M31, M101, M83, and M51. 
In fact, with typical values of $N_H\sim 10^{21}$~cm$^{-2}$,
the {\it Chandra} exposures we analyzed should allow us to detect
and identify all NBWDs 
with $M_{WD}$ larger than
$0.8-0.9\, M_\odot$ ($1.2-1.3\, M_\odot$) 
in M31 (M81 and M83).  
Elliptical galaxies are good places in which to hunt for bright SSSs 
because they have relatively small amounts of gas and dust.

If, therefore, (1) the class of 
Type Ia progenitors consists primarily of accreting white dwarfs
with mass close to $M_C,$ and (2) these white dwarfs 
are burning the accreted matter
in a quasi-steady manner, and (3) this produces luminosities and
temperatures in the expected regime, then these galaxies should each contain
at least several hundred hot, bright, and detectable SSSs.
   
\subsection{Observations}
\begin{table}
 \centering
\caption{Soft Sources in External Galaxies}
\label{tlab}
\begin{tabular}{cccc}\hline
Galaxy  & SSSs & QSSs & Other Sources \\
\hline
M101       & 42  & 21  & 24   \\
M83        & 28  & 26  & 74   \\
M51        & 15  & 21  & 56   \\
M104       & 5   & 17  & 100  \\
NGC4472    & 5   & 22  & 184  \\
NGC4697    & 4   & 15  & 72   \\
\hline
\end{tabular}
\end{table}

\subsubsection{Individual Galaxies} 
Table 1 shows the numbers of soft x-ray sources in each of six
galaxies that have been carefully studied (Di\thinspace Stefano
et al.\, 2003a, 2003b, 2004a, 2004b).
M101, M83, and M51 are spiral galaxies. M104 is a bulge-dominated spiral,
while both NGC4472 and NGC4697 are elliptical galaxies. 
Two categories of soft sources are listed. 

(1) {\bf SSSs} refer to sources with
values of $k\, T$ below $150$~eV. These should encompass the majority
of NBWDs, including all of the Type Ia progenitors. 
In each case, the numbers of SSSs actually detected are smaller than
the numbers of predicted Type Ia 
progenitors by more than an order of magnitude.
Furthermore, 
only a fraction of the detected SSSs are candidate \pr s of \t1e.  
This is because 
luminosities and temperatures of many of the detected SSSs are so low that,
if they correspond to NBWDs, the mass of the white dwarf is 
much smaller than $M_C$ (see, e.g., Pence et al. 2001). 
On the other hand, some SSSs are 
too luminous to correspond to NBWDs. 
Some ultraluminous SSSs, 
with 
x-ray luminosities above $10^{39}$~erg~s$^{-1}$, 
may be accreting black holes (Kong \& Di~Stefano 2005; 
Mukai et al. 2005;
Kong et al. 2004). 

(2)~{\bf Quasisoft sources (QSSs)} are 
sources that are slightly harder than    
SSSs, but which exhibit little or no emission above $2$~keV
(Di~Stefano \& Kong 2004; Di~Stefano et al. 2004a; Di~Stefano et al. 2004b). 
With effective values of $k\, T$ above $\sim 150$~eV,
some could be NBWDs with mass near $M_C:$ the addition of a
small hard component and/or high levels of absorption could  
cause such a source to have a QSS spectrum. Nevertheless, many QSSs
are too hard to be associated with NBWDS.
At most a small fraction could be progenitors of Type Ia SNe. 

{\bf Results:} 
At most, a modest fraction of all SSSs and QSSs
could be \pr s of \t1e. Yet, even if the total of all SSSs {\it and}
QSSs in each galaxy were \pr s, they would constitute less than a tenth
of the total NBWDs with mass near $M_C$ needed to provide the
observed rate of \t1e. 
Therefore, even if some of the detected SSSs and QSS are 
\pr s, there is a clear and large mismatch, with at least $90-99\%$ 
of the 
actively accreting \pr s not appearing as SSSs or QSSs.
Figure~2 also indicates that the observations already place limits on the
numbers of lower-mass NBWDs in each of several galaxies.

It may be important to note, however, that there is an interesting trend
in the data. That is, both the rate of \t1e and
the number of SSSs are significantly larger in late-type
galaxies than in early-type galaxies.

If the rate of
mass transfer is high enough to support quasisteady-nuclear burning, then
the character of the disk is remarkable, because the total accretion energy
is only a small fraction of the energy provided to the disk by 
the \wdf\ through irradiation and heating. The inner disk itself becomes a
source of supersoft x-rays. At x-ray wavelengths, the effect is therefore to 
make the source marginally brighter, but to likely keep it in the
SSS regime. Should it b 
 
\subsubsection{A Large Sample of Galaxies:} 
An automated source detection and identification process was employed
by Liu (2008) to study x-ray sources in fields containing
383 nearby galaxies. SSSs and QSSs were identified using the
same algorithm used for the galaxies in Table 1.
Liu found that only $2.6\%$ of all sources
bright enough for spectral classification
are SSS. For every SSS there are roughly four QSSs.\footnote{The high ratio of
QSSs to SSSs may reflect the prevalence of 
older stellar populations among the galaxies
in Liu's survey.} 
The combination of SSSs and
QSSs constitute about $13\%$ of all x-ray sources.
Yet, the total numbers of x-ray sources per 
galaxy were generally several times smaller
than the expected population of nuclear-burning Type Ia progenitors.
Thus, the discrepancy between the numbers
of progenitors and the total numbers of soft sources per galaxy
is typically larger than an order of magnitude.
Furthermore, a significant fraction of the soft sources have
x-ray luminosities above $10^{39}$~erg~s$^{-1}$, and are therefore
not good candidates for white dwarf accretors. 
  
Liu created merged images to achieve effective exposure times far longer
than those utilized in Figure 2. The total exposure time
of M101, for example, is roughly 
ten times the value used to create Figure~2. This means that Liu's
catalog can be used to place stringent limits on sub-Chandrasekhar
models (with white dwarf masses as low as $0.8-0.9\, M_\odot$) 
 as well as on Chandrasekhar-mass models.

Note that, in addition to the observations of the previous subsection,
and the work of Liu (2008), many studies of 
extragalactic x-ray sources  have identified SSSs. See, e.g.,
Sarazin et al.(2001);  
Swartz et al.(2002); 
Pence et al.(2001); 
Jenkins et al.(2004);  
Fabbiano et al.(2003). 
 We
focus on the Liu (2008) paper because it is comprehensive
and uses uniform detection and selection procedures. 
Note, however, that the results of the individual studies are
consistent with those of Liu: the total numbers of bright SSSs
per galaxy is small, consistent with the source counts presented in Table 1.
  
\subsubsection{M31}
M31 
is a special case because of its proximity. 
A source with x-ray luminosity ($0.2-10$ keV) of 
$10^{38} {\rm erg~s}^{-1}$, possibly corresponding to a NBWD
with mass within roughly $0.15\, M_\odot$ of $M_C,$ 
would produce a count rate of $0.04\ {\rm s}^{-1}$ in ACIS-S,
or $0.014\ {\rm s}^{-1}$ in ACIS-I. 
Since most exposures of M31 are longer than $5$~ksec,
these sources are readily detected 
and their spectra can be classified as SSS. A source 
which exhibits a small hard component  
might appear as a QSS, producing an even higher count rate. 
Di\thinspace Stefano et al.\, (2004b) 
published a list of $42$ SSSs and QSSs drawn from
three regions in the disk and also from the galactic bulge. 
Of these, only $4$ sources
($3$ of them in the bulge), had count rates above 
$0.01\, {s}^{-1}$.
Most of the detected soft sources are too dim to be high-mass NBWDs. 
These observations cover only a fraction of the galaxy. 
(Note that most of the SSSs
are in the center, which was well covered, so the total number of sources
we expect does not scale with the area covered.)  
Orio (2006) 
identified SSSs and QSSs in M31 by using deep 
{\it XMM-Newton} observations that covered the entire body of the 
galaxy. Using x-ray spectral information alone, Orio
found significant contamination from supernova remnants. 
The data contained evidence for $15$
SSSs and $18$ QSSs that were not associated with known 
supernova remnants.
 In addition, Orio found that
about one in ten nova  outbursts in M31 produces SSS-behavior post nova.
The link to novae has been studied in 
more detail using additional data sets
(Pietsch et al.\, 2006; 
Henze et al.\, 2009).
della Valle
\& Livio (1996) have demonstrated that only $\sim 10\%$ of all 
novae are likely to be
RNe, retaining most of the matter they accrete.  
Most novae, therefore, are not \t1\ \pr\ candidates.
In summary, the total number of SSSs detected in M31 is smaller
than the requisite number of \pr s, and many of the ones that are
   detected are not good candidates for Type~Ia \pr s.  

In fact in M31, where typical values of $N_H$ are $\sim 10^{21}$~cm$^{-2},$
we likely have a complete census of NBWDs with mass greater than roughly
$0.9\, M_\odot.$ This places strong limits on sub-Chandrasekhar models. 
 
\subsubsection{The Milky Way}
Only a handful of SSSs have been discovered in the Milky
Way. This small number is consistent with the fact that
absorption by the interstellar medium (ISM) in the \Gc\ disk can \obe\ the soft \rdtn\
emanating from \l-m\ \N s. If, however, the Galaxy contains
$500-1000$~\h-m\ \N s, $5-10$ of them should lie
within roughly a kpc. If these have values of $k\, T$ 
in the range $75-100$~eV and luminosities of $\sim 10^{38}$~erg~s$^{-1}$,
they will be easily detected by ROSAT, {\it Chandra}, and
{\it XMM-Newton}.

{For {\it Chandra} count rates we can scale from
the numbers given in the paragraph on M31; the \xmm\ count rates would
be $\sim 4$~times higher. While the 
\ro\ count rates would be smaller than the \ch\ rates by a factor
of $\sim 2$, the {\it All-Sky Survey} could have discovered all
\h-m\ \N s within a kpc, even along directions in the sky in which
the exposure times were shortest. The persistent SSSs 
discovered by {\it ROSAT}
were dimmer and cooler than \h-m\ 
NBWDs (Greiner 2000). 
}  
The failure to detect bright, hot SSSs and/or QSSs in the neighborhood
($< 1$~kpc) of the Sun is an independent indication that such sources do
not comprise the majority of \t1\ \pr s. 

\subsubsection{Pre- and Post-explosion Comparisons}
The most direct way to
determine if \t1e are generated by SSSs is to check pre-explosion
x-ray images. Voss \& 
Nelemans (2008) found $4$ galaxies that had been observed by \ch\
prior to a \t1\ \exn.    
There was no sign of an x-ray source in the vicinities of $3$ of the
supernovae. 
The detection limits were below those expected for high-mass NBWDs.
In one case, there was a nearby source, but it was
$1.15^{\prime\prime}$ from the supernova ($\sim 80$~pc in projection; 
Roelofs et al.\, 2008), 
and was neither an SSS nor 
a QSS.

\begin{figure*}
\begin{center}
\psfig{file=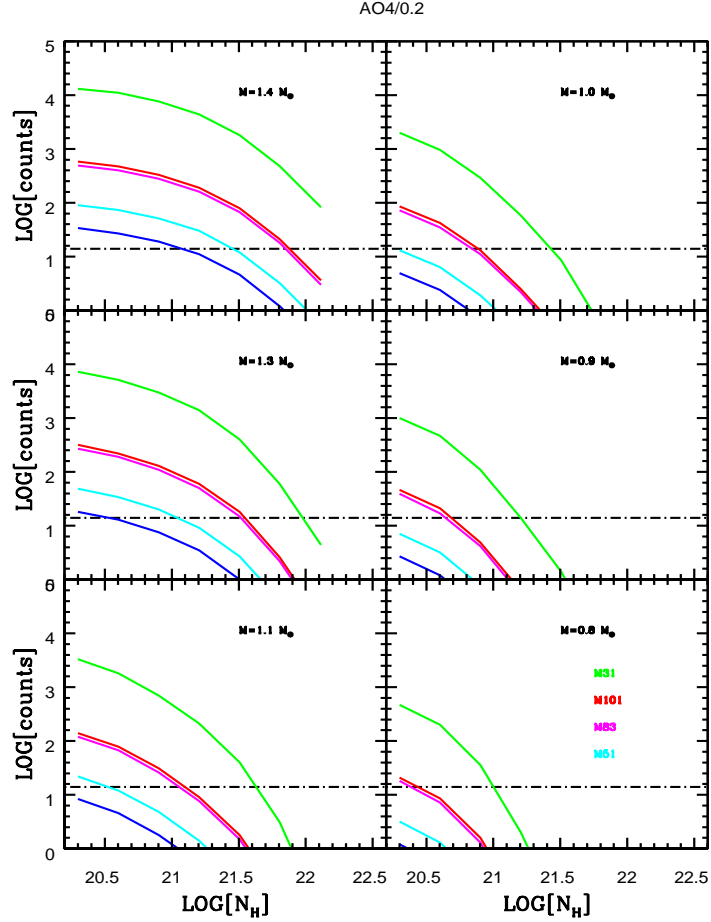,
height=5.0in,width=4.0in,angle=-0.0}
\vspace{.8 true in}
\caption{
Each panel refers to a quasi-steady
 NBWD with the listed mass; $L$ and $T$ were
taken from Iben (1982).
Within each panel, each
curve corresponds to a single galaxy (M31, green; M101, red; M83, magenta;
M51, cyan; NGC 4697, blue); the galaxy's distance and the longest ACIS-S
exposure we have analyzed were used to compute the numbers of
counts. We assumed the sensitivity of {\it Chandra's} ACIS-S
as computed by PIMMS for AO4.
The horizontal dot-dashed line shown on each panel corresponds
to $14$ counts, the number needed to ascertain whether a given
source can be classified as an SSS (see Di\thinspace Stefano \& Kong 2003).
}
\end{center}
\end{figure*}

\section{Conclusions}

\subsection{SSSs for the Long Term?} 
The discovery of SSSs seemed to provide hope that Type~Ia progenitors
could be identified in a fairly straightforward way, through the 
signatures of their soft x-ray emission.
While some bright SSSs may be progenitors of Type~Ia
supernovae, we can now say conclusively that the majority of
the progenitors do not appear as bright SSSs during intervals long
enough ($\sim 10^5$~yrs) to allow quasisteady burning of the necessary
amounts of accreting matter.  
The discrepancy is at least an order of magnitude, and may
be as much as two orders of magnitude.

In addition, we find that existing data already place restrictions on 
sub-Chandrasekhar
models (Figure 2). 
These restrictions will be tightened as more data are analyzed, and additional
exposures with {\it Chandra} and {\it XMM-Newton} become available.

\subsection{SSSs for the Short Term?} 

Although we have ruled out the possibility that the majority of 
\t1\ \pr s appear as SSSs during much of the all-important 
mass-gaining phase, there is almost certainly a tight link between SSSs 
and \t1\ \pr s.  
While there are several ways to ``hide'' the soft x-ray emission from
NBWDs (see the companion paper, Di\thinspace Stefano 2009),
 it is difficult to construct a model in which a \wdf\ burns 
a tenth of a solar mass or more, yet never
exhibits at least brief SSS or QSS phases.
For example, some \pr s may 
experience short periods of SSS behavior during the time when
the system approaches or passes the minimum accretion rate
needed for steady burning. 
This can happen both as mass transfer ramps up
and also as it ramps down.  
In addition, a small fraction of 
binaries in which the \wdf\ reaches $M_C$ 
 can sustain mass transfer rates in the steady burning regime for  
most of the mass-transfer phase. (See, e.g., van den Heuvel et al. 1992; Rappaport,  
 Di~Stefano \& Smith 1994; Di~Stefano \& Nelson 1996; Yungelson et al. 1996; L{\"u} et al. 2006;
Han
\& Podsiadlowski 2006.)

It is therefore possible that a portion of the observed SSSs and QSSs 
are  
CO \wdf s on their way to the Chandrasekhar mass.  
It is important to identify candidates. 
Given the expected temperature range, the NBWDs that are
\t1\ \pr s  could appear as very hot SSSs. If there is a small
hard component and/or if absorption is important, they could instead
appear as the coolest of QSSs. 
Wherever possible, these candidates should be monitored.
Those that 
are not high-mass NBWDs will be interesting anyway, because 
they are unusual states of neutron-star or black-hole accretors.
For example, some could be black holes of $\sim 50-100\, M_\odot$
in thermal dominant states Di~Stefano et al.\, 2004a).

The candidates that turn out to be NBWDs could be near   
explosion. The likelihood that an individual source
is about to ``go off'' is small.
Even a \wdf\ within $0.01\, M_\odot$ may take more than $1000$
years to achieve $M_C$. Nevertheless,  
 by identifying a large pool
of candidates, drawn from the hundreds of galaxies 
observed with {\it Chandra} and {\it XMM-Newton}, we increase the
chance of having a useful pre-explosion image of a \t1\ that
explodes within the next decade.\footnote{In the companion paper we also discuss 
the discovery of \pr s at other wavelengths (IR, optical, and UV) in the context of 
binary models.}       
  
\subsection{A Multiwavelengh Guide}

The restrictions on the 
numbers of SSSs we have derived above do not necessarily translate into
restrictions on the numbers of NBWDs. 
In fact, theoretical considerations predict the presence of a population of
 high-mass NBWDs. First, starting with a large population of binaries,
a subset will pass through a phase in which a high-mass \wdf\ accretes matter
at a high rate (see, e.g., Rappaport,  Di~Stefano \& Smith 1994; Yungelson et al. 1996). 
Second, we know that \t1e occur, and that every
evolutionary pathway
to explosion passes through channels that involve high accretion rates onto 
a \wdf\ with mass larger than $\sim 0.8\, M_\odot.$ Certainly
a long phase of nuclear-burning is required in \s-d\ models (Equation 3).
This would be true in sub-Chandrasekhar models as well.
Furthermore, we show in the companion paper that even many \d-d\ \bis\
require a prior \s-d\ phase during which nuclear burning is expected (Di~Stefano 2009).
 
We have also found that the numbers of NBWDs needed per galaxy is large.
Thus, at any given time, galaxies such as the Milky Way house 
$\sim 1000$ NBWDs with masses near $M_C.$ In sub-Chandrasekhar models, or
\d-d\ models the average mass of the NBWDs may be smaller, but 
nevertheless larger than $0.8\, M_\odot.$ 
This increases our chances of establishing the existence of a population
of \pr s in M31 and in other nearby galaxies. Perhaps most exciting, the
large number guarantees that some \pr s that are actively accreting 
NBWDs must lie within a kpc of Earth. If we can identify these, we
will be able to make detailed studies of the \pr s and the processes
that allow high-mass \wdf s to grow in mass.  

In fact, we have almost certainly detected NBWDs that are \t1\ \pr s, in the Milky Way
and in nearby galaxies. These must be bright systems. Since the bulk of the energy produced  
by nuclear burning does not escape at x-ray wavelengths, it must be emitted
in other portions of the electromagnetic spectrum.  
Radiation from the \wdf\ could, for example, be reprocessed by 
winds from the \bi . In this case, the spectrum would be shifted toward
longer wavelengths, and the \pr s could be highly luminous in the IR.
Alternatively, if the photospheres of massive NBWDs tend to be large,
compared with size of the \wdf , the
power can be
emitted primarily in the UV. If \pr s pass through phases of SSS behavior,
even with a low duty cycle, they may be associated with large, distinctive
ionization nebulae (Rappaport et al. 1994; Chiang \& Rappaport 1996; Di~Stefano,
Paerels \& Rappaport 1995).   
In the companion paper (Di~Stefano 2009) we link the possible
signatures of nuclear burning to the characteristics of the binary
models for \t1\ \pr s.  

\medskip

\noindent{\bf Summary:} Nuclear burning appears to be required in order for Type~Ia
supernovae to occur. As the companion paper shows, even many 
\d-d\ \pr s of \t1e 
are expected to pass through an earlier \s-d\ phase in   
which nuclear burning occurs. 
We have demonstrated that the majority of the \pr s do not    
appear as SSSs or QSSs during most of the all-important nuclear-burning phase. 
The energy must escape in other wavebands, making it possible to detect and
identify dozens of \pr s in our own Galaxy, and comparable or perhaps even larger
numbers in M31 and other nearby galaxies. 
Given the long duration of the requisite nuclear-burning phases, it would be surprising if
some \pr s do not appear as SSSs or QSSs, at least during part of their evolution.
It is therefore still interesting to explore the connection between soft x-ray sources
and \t1\ \pr s, even as we extend searches to bright IR and UV sources, and large
ionization nebulae, so as to find evidence of the centrally-important nuclear-burning phase. 
The bottom line is that, although \t1e are rare and therefore tend to occur at great distance from us,
thousands of the \pr s are nearby, and 
are luminous enough that 
we can study them in detail well before
they explode.

\bigskip

\noindent{\bf Acknowledgements:} It is a pleasure to acknowledge      
helpful conversations and 
comments from many of the participants
(especially Ed van den Heuvel, Lev Yungelson, and Jim Liebert), 
 of the KITP conference
and workshop on {\it Accretion and Explosion} held at UC Santa Barbara in
2007.
This work was supported in part by an LTSA grant from NASA
and by funding from the Smithsonian Institution.


\bigskip

\begin{thebibliography}{}


\bibitem[Branch et al.(1995)]{1995PASP..107.1019B} Branch, D., Livio, M., 
Yungelson, L.~R., Boffi, F.~R., \& Baron, E.\ 1995, \pasp, 107, 1019 

\bibitem[Cappellaro et 
al.(1993)]{1993A&A...273..383C} Cappellaro, E., 
Turatto, M., Benetti, S., Tsvetkov, D.~Y., Bartunov, O.~S., \& 
Makarova, I.~N.\ 1993, \aap, 273, 383 

\bibitem[Catal{\'a}n et al.(2009)]{2009JPhCS.172a2007C} Catal{\'a}n, S., 
Isern, J., Garc{\'{\i}}a-Berro, E., 
\& Ribas, I.\ 2009, Journal of Physics Conference Series, 172, 012007 


\bibitem[Chiang
\& Rappaport(1996)]{} Chiang, E., \& Rappaport, S.\ 1996, \apj, 469, 255

\bibitem[della Valle 
\& Livio(1996)]{1996ApJ...473..240D} della Valle, M., \& Livio, M.\ 1996, \apj, 473, 240 

\bibitem[Dilday et al.(2008)]{2008ApJ...682..262D} Dilday, B., et al.\ 
2008, \apj, 682, 262 



\bibitem[Di Stefano (2009)]{2009} Di Stefano, R., 2009, 
{\sl The progenitors of Type Ia Supernovae: II.~A~Guide~to~Discovery}, in preparation.

\bibitem[Di Stefano et al.(2009)]{2009arXiv0909.2046D} Di Stefano, R., 
Primini, F.~A., Liu, J., Kong, A., \& Patel, B.\ 2009, arXiv:0909.2046 

\bibitem[Di Stefano et al.(2004)]{} Di Stefano, R.,
Primini, F.~A., Kong, A.~K.~H., \& Russo, T.\ 2004a, arXiv:astro-ph/0405238

\bibitem[Di Stefano et al.(2004)]{} Di Stefano, R., et
al.\ 2004b, \apj, 610, 247

\bibitem[Di Stefano 
\& Kong(2004)]{2004ApJ...609..710D} Di Stefano, R., \& Kong, A.~K.~H.\ 2004, \apj, 609, 710
 
\bibitem[Di Stefano et al.(2003)]{} Di Stefano, R.,
Friedman, R., Kundu, A., \& Kong, A.~K.~H.\ 2003a, arXiv:astro-ph/0312391

\bibitem[Di Stefano et al.(2003)]{} Di Stefano, R.
et al.\, 2003b, \apj, 599, 1067

\bibitem[Di Stefano
\& Kong(2003)]{} Di Stefano, R., \& Kong, A.~K.~H.\ 2003b, \apj, 592, 884


\bibitem[Di Stefano 
\& Nelson(1996)]{1996LNP...472....3D} Di Stefano, R., \& Nelson, L.~A.\ 1996, Supersoft X-Ray Sources, 472, 3 

\bibitem[Di Stefano et al.(1995)]{1995ApJ...450..705D} Di Stefano, R., 
Paerels, F., \& Rappaport, S.\ 1995, \apj, 450, 705 

\bibitem[Di Stefano
\& Rappaport(1994)]{} Di Stefano, R., \& Rappaport, S.\ 1994, \apj, 437, 733

\bibitem[Dobbie et al.(2006)]{2006MNRAS.369..383D} Dobbie, P.~D., et al.\ 
2006, \mnras, 369, 383 

\bibitem[Fabbiano et al.(2003)]{2003ApJ...591..843F} Fabbiano, G., King, 
A.~R., Zezas, A., Ponman, T.~J., Rots, A., 
\& Schweizer, F.\ 2003, \apj, 591, 843 


\bibitem[Fujimoto(1982)]{1982ApJ...257..767F} Fujimoto, M.~Y.\ 1982, \apj, 
257, 767 

\bibitem[Graham et al.(2008)]{2008AJ....135.1343G} Graham, M.~L., et al.\ 
2008, \aj, 135, 1343 



\bibitem[Greiner(2000)]{2000NewA....5..137G} Greiner, J.\ 2000, New 
Astronomy, 5, 137 

\bibitem[Han 
\& Podsiadlowski(2006)]{2006MNRAS.368.1095H} Han, Z., \& Podsiadlowski, P.\ 2006, \mnras, 368, 1095 

\bibitem[Henze et 
al.(2009)]{2009A&A...500..769H} Henze, M., et al.\ 2009, \aap, 500, 769 


\bibitem[Iben(1982)]{1982ApJ...259..244I} Iben, I., Jr.\ 1982, \apj, 259, 
244 


\bibitem[Jenkins et al.(2005)]{2005MNRAS.357..401J} Jenkins, L.~P., 
Roberts, T.~P., Warwick, R.~S., Kilgard, R.~E., 
\& Ward, M.~J.\ 2005, \mnras, 357, 401 

\bibitem[Kalirai et al.(2008)]{2008ApJ...676..594K} Kalirai, J.~S., Hansen, 
B.~M.~S., Kelson, D.~D., Reitzel, D.~B., Rich, R.~M., 
\& Richer, H.~B.\ 2008, \apj, 676, 594 

\bibitem[Kong
\& Di Stefano(2005)]{} Kong, A.~K.~H., \& Di Stefano, R.\ 2005, 
\apjl, 632, L107 

\bibitem[Kong et al.(2004)]{2004ApJ...617L..49K} Kong, A.~K.~H., 
Di~Stefano, R., \& Yuan, F.\ 2004, \apjl, 617, L49 

\bibitem[Kong 
\& Di Stefano(2003)]{2003ApJ...590L..13K} Kong, A.~K.~H., \& Di Stefano, R.\ 2003, \apjl, 590, L13 


\bibitem[Kotak(2009)]{2009ASPC..401..150K} Kotak, R.\ 2009, Astronomical 
Society of the Pacific Conference Series, 401, 150 

\bibitem[Kuznetsova et al.(2008)]{2008ApJ...673..981K} Kuznetsova, N., et 
al.\ 2008, \apj, 673, 981 




\bibitem[Liu(2008)]{} Liu, J.\ 2008, arXiv:0811.0804

\bibitem[L{\"u} et al.(2006)]{2006MNRAS.372.1389L} L{\"u}, G., Yungelson, 
L., \& Han, Z.\ 2006, \mnras, 372, 1389 


\bibitem[Miller 
\& Scalo(1979)]{1979ApJS...41..513M} Miller, G.~E., \& Scalo, J.~M.\ 1979, \apjs, 41, 513 

\bibitem[Mukai et al.(2005)]{} Mukai, K.
et al.\, 2005, \apj, 634, 1085

\bibitem[Nomoto(1982)]{1982ApJ...253..798N} Nomoto, K.\ 1982, \apj, 253, 
798 


\bibitem[Nomoto et al.(1979)]{1979wdvd.coll...56N} Nomoto, K., Miyaji, S., 
Sugimoto, D., 
\& Yokoi, K.\ 1979, IAU Colloq.~53: White Dwarfs and Variable Degenerate Stars, 56 



\bibitem[Orio(2006)]{} Orio, M.\ 2006, \apj, 643, 844


\bibitem[Paczy{\'n}ski(1971)]{1971ARA&A...9..183P} Paczy{\'n}ski, B.\ 1971, \araa, 9, 183 



\bibitem[Panagia et al.(2007)]{2007AIPC..924..373P} Panagia, N., Della 
Valle, M., 
\& Mannucci, F.\ 2007, The Multicolored Landscape of Compact Objects and Their Explosive Origins, 924, 373 

\bibitem[Pence et al.(2001)]{2001ApJ...561..189P} Pence, W.~D., Snowden, 
S.~L., Mukai, K., \& Kuntz, K.~D.\ 2001, \apj, 561, 189 



\bibitem[Pietsch et
al.(2005)]{} Pietsch, W. et al.\,
 2005, A\& A , 442, 879

\bibitem[Poznanski et al.(2007)]{2007MNRAS.382.1169P} Poznanski, D., et 
al.\ 2007, \mnras, 382, 1169 



\bibitem[Rappaport et al.(1994)]{} Rappaport, S.,
Di~Stefano, R., \& Smith, J.~D.\ 1994, \apj, 426, 692

\bibitem[Rappaport et al.(1994)]{1994ApJ...431..237R} Rappaport, S., 
Chiang, E., Kallman, T., \& Malina, R.\ 1994, \apj, 431, 237


\bibitem[Riess et al.(2007)]{2007ApJ...659...98R} Riess, A.~G., et al.\ 
2007, \apj, 659, 98 

\bibitem[Roelofs et al.(2008)]{2008MNRAS.391..290R} Roelofs, G., Bassa, C., 
Voss, R., \& Nelemans, G.\ 2008, \mnras, 391, 290 

\bibitem[Sarazin et al.(2001)]{2001ApJ...556..533S} Sarazin, C.~L., Irwin, 
J.~A., \& Bregman, J.~N.\ 2001, \apj, 556, 533 

\bibitem[Shen 
\& Bildsten(2008)]{2008ApJ...678.1530S} Shen, K.~J., \& Bildsten, L.\ 2008, \apj, 678, 1530 


\bibitem[Salpeter(1955)]{1955ApJ...121..161S} Salpeter, E.~E.\ 1955, \apj, 
121, 161 

\bibitem[Swartz et al.(2002)]{2002ApJ...574..382S} Swartz, D.~A., Ghosh, 
K.~K., Suleimanov, V., Tennant, A.~F., \& Wu, K.\ 2002, \apj, 574, 382 

\bibitem[Turatto et al.(1994)]{1994AJ....108..202T} Turatto, M., 
Cappellaro, E., \& Benetti, S.\ 1994, \aj, 108, 202 

\bibitem[]{}
van den Heuvel, E.P.J., Bhattacharya, D., Nomoto, K., \& Rappaport, S.A.
1992, A\&A, 262, 97.


\bibitem[Voss 
\& Nelemans(2008)]{2008Natur.451..802V} Voss, R., \& Nelemans, G.\ 2008, \nat, 451, 802 

\bibitem[Weidemann 
\& Koester(1983)]{1983A&A...121...77W} Weidemann, V., \& Koester, D.\ 1983, \aap, 121, 77 

\bibitem[Williams(2007)]{2007ASPC..372...85W} Williams, K.~A.\ 2007, 15th 
European Workshop on White Dwarfs, 372, 85 


\end{thebibliography}
\end{document}